\documentclass[aps,prc,twocolumn,superscriptaddress]{revtex4-1}
\usepackage[utf8]{inputenc}
\usepackage[english]{babel}
\usepackage{bm}
\usepackage{graphicx}
\usepackage{epstopdf}
\usepackage{amsmath}
\usepackage{amssymb}
\usepackage{booktabs}
\usepackage{enumitem}
\usepackage{color}
\usepackage[caption=false]{subfig}
\usepackage [autostyle, english = american]{csquotes}
\MakeOuterQuote{"}

\usepackage{mathtools} 	
\usepackage{comment} 	
\usepackage{multirow}	
\usepackage{gensymb}
\usepackage{epstopdf}
\usepackage[]{siunitx}  
\DeclareSIUnit\sample{S}
\DeclareSIUnit\bits{bits}

\usepackage{lineno} 	
\newcommand*\patchAmsMathEnvironmentForLineno[1]{%
  \expandafter\let\csname old#1\expandafter\endcsname\csname #1\endcsname
  \expandafter\let\csname oldend#1\expandafter\endcsname\csname end#1\endcsname
  \renewenvironment{#1}%
     {\linenomath\csname old#1\endcsname}%
     {\csname oldend#1\endcsname\endlinenomath}}%
\newcommand*\patchBothAmsMathEnvironmentsForLineno[1]{%
  \patchAmsMathEnvironmentForLineno{#1}%
  \patchAmsMathEnvironmentForLineno{#1*}}%
\AtBeginDocument{%
\patchBothAmsMathEnvironmentsForLineno{equation}%
\patchBothAmsMathEnvironmentsForLineno{align}%
\patchBothAmsMathEnvironmentsForLineno{flalign}%
\patchBothAmsMathEnvironmentsForLineno{alignat}%
\patchBothAmsMathEnvironmentsForLineno{gather}%
\patchBothAmsMathEnvironmentsForLineno{multline}%
}
\usepackage{xspace} 	

\newcommand{\znbb} {$0\nu\!\beta\!\beta$\xspace}

\newcommand{\Efield} {380~V/cm\xspace}
\newcommand{\EfieldNoUnit} {380\xspace}
\newcommand{\CathToV} {192.4~mm\xspace}
\newcommand{\UToV} {6~mm\xspace}

\makeatletter
\renewcommand\tableofcontents{%
    \@starttoc{toc}%
}
\makeatother

\newif\ifshowchanges
\showchangesfalse

\newcommand{\vo}[1]{}

\ifshowchanges
\usepackage{soul}
\soulregister\cite7
\soulregister\ref7
	\renewcommand{\vo}[1]{{\textcolor{red}{\st{#1}}}}
    
\fi

\begin{document}

\title{Measurement of the Drift Velocity and Transverse Diffusion of Electrons in Liquid Xenon with the EXO-200 Detector}
\collaboration{EXO-200 Collaboration}

\newcommand{\Alabama}{\affiliation{Department of Physics and Astronomy, University of Alabama, Tuscaloosa, Alabama 35487, USA}}
\newcommand{\Alberta}{\affiliation{University of Alberta, Edmonton, Alberta, Canada}}
\newcommand{\Bern}{\affiliation{LHEP, Albert Einstein Center, University of Bern, Bern, Switzerland}}
\newcommand{\CALTECH}{\affiliation{Kellogg Lab, Caltech, Pasadena, California 91125, USA}}
\newcommand{\Carleton}{\affiliation{Physics Department, Carleton University, Ottawa, Ontario K1S 5B6, Canada}}
\newcommand{\CSU}{\affiliation{Physics Department, Colorado State University, Fort Collins, Colorado 80523, USA}}
\newcommand{\Drexel}{\affiliation{Department of Physics, Drexel University, Philadelphia, Pennsylvania 19104, USA}}
\newcommand{\Duke}{\affiliation{Department of Physics, Duke University, and Triangle Universities Nuclear Laboratory (TUNL), Durham, North Carolina 27708, USA}}
\newcommand{\IBS}{\affiliation{IBS Center for Underground Physics, Daejeon, Korea}}
\newcommand{\IHEP}{\affiliation{Institute of High Energy Physics, Beijing, China}}
\newcommand{\Illinois}{\affiliation{Physics Department, University of Illinois, Urbana-Champaign, Illinois 61801, USA}}
\newcommand{\Indiana}{\affiliation{Physics Department and CEEM, Indiana University, Bloomington, Indiana 47405, USA}}
\newcommand{\ITEP}{\affiliation{Institute for Theoretical and Experimental Physics, Moscow, Russia}}
\newcommand{\Laurentian}{\affiliation{Department of Physics, Laurentian University, Sudbury, Ontario P3E 2C6, Canada}}
\newcommand{\Maryland}{\affiliation{Physics Department, University of Maryland, College Park, Maryland 20742, USA}}
\newcommand{\McGill}{\affiliation{Physics Department, McGill University, Montreal, Quebec H3A 2T8, Canada}}
\newcommand{\Munich}{\affiliation{Technische Universit\"at M\"unchen, Physikdepartment and Excellence Cluster Universe, Garching 80805, Germany}}
\newcommand{\SDakota}{\affiliation{Physics Department, University of South Dakota, Vermillion, South Dakota 57069, USA}}
\newcommand{\Seoul}{\affiliation{Department of Physics, University of Seoul, Seoul, Korea}}
\newcommand{\SLAC}{\affiliation{SLAC National Accelerator Laboratory, Menlo Park, California 94025, USA}}
\newcommand{\Stanford}{\affiliation{Physics Department, Stanford University, Stanford, California 94305, USA}}
\newcommand{\Stony}{\affiliation{Department of Physics and Astronomy, Stony Brook University, SUNY, Stony Brook, New York 11794, USA}}
\newcommand{\TRIUMF}{\affiliation{TRIUMF, Vancouver, British Columbia V6T 2A3, Canada}}
\newcommand{\UMass}{\affiliation{Amherst Center for Fundamental Interactions and Physics Department, University of Massachusetts, Amherst, MA 01003, USA}}
\newcommand{\WIPP}{\affiliation{Waste Isolation Pilot Plant, Carlsbad, New Mexico 88220, USA}}
\newcommand{\SNOLAB}{\affiliation{SNOLAB, Sudbury, Ontario P3Y 1N2, Canada}}
\newcommand{\Yale}{\affiliation{Department of Physics, Yale University, New Haven, Connecticut 06511, USA}}

\author{J.B.~Albert}\Indiana
\author{P.S.~Barbeau}\Duke
\author{D.~Beck}\Illinois
\author{V.~Belov}\ITEP
\author{M.~Breidenbach}\SLAC
\author{T.~Brunner}\McGill \TRIUMF
\author{A.~Burenkov}\ITEP
\author{G.F.~Cao}\IHEP
\author{W.R.~Cen}\IHEP
\author{C.~Chambers}\CSU
\author{B.~Cleveland}\Laurentian \SNOLAB
\author{M.~Coon}\Illinois
\author{A.~Craycraft}\CSU
\author{T.~Daniels}\SLAC
\author{M.~Danilov}\ITEP
\author{S.J.~Daugherty}\Indiana
\author{J.~Daughhetee}\SDakota
\author{J.~Davis}\SLAC
\author{S.~Delaquis}\SLAC
\author{A.~Der Mesrobian-Kabakian}\Laurentian
\author{R.~DeVoe}\Stanford
\author{T.~Didberidze}\Alabama
\author{J.~Dilling}\TRIUMF
\author{A.~Dolgolenko}\ITEP
\author{M.J.~Dolinski}\Drexel
\author{M.~Dunford}\Carleton
\author{W.~Fairbank Jr.}\CSU
\author{J.~Farine}\Laurentian
\author{S.~Feyzbakhsh}\UMass 	
\author{P.~Fierlinger}\Munich
\author{D.~Fudenberg}\Stanford
\author{R.~Gornea}\Carleton \TRIUMF
\author{K.~Graham}\Carleton
\author{G.~Gratta}\Stanford
\author{C.~Hall}\Maryland
\author{M.~Hughes}\Alabama
\author{M.J.~Jewell}\Stanford
\author{A.~Johnson}\SLAC
\author{T.N.~Johnson}\Indiana
\author{S.~Johnston}\UMass
\author{A.~Karelin}\ITEP
\author{L.J.~Kaufman}\Indiana
\author{R.~Killick}\Carleton
\author{T.~Koffas}\Carleton
\author{S.~Kravitz}\Stanford
\author{R.~Kr\"{u}cken}\TRIUMF
\author{A.~Kuchenkov}\ITEP
\author{K.S.~Kumar}\Stony
\author{Y.~Lan}\TRIUMF
\author{D.S.~Leonard}\IBS
\author{C.~Licciardi}\Carleton
\author{Y.H.~Lin}\Drexel
\author{R.~MacLellan}\SDakota
\author{M.G.~Marino}\Munich
\author{B.~Mong}\SLAC
\author{D.~Moore}\Yale
\author{O.~Njoya}\Stony
\author{R.~Nelson}\WIPP
\author{A.~Odian}\SLAC
\author{I.~Ostrovskiy}\Alabama
\author{A.~Piepke}\Alabama
\author{A.~Pocar}\UMass
\author{C.Y.~Prescott}\SLAC
\author{F.~Reti\`{e}re}\TRIUMF
\author{P.C.~Rowson}\SLAC
\author{J.J.~Russell}\SLAC
\author{A.~Schubert}\Stanford
\author{D.~Sinclair}\Carleton \TRIUMF
\author{E.~Smith}\Drexel
\author{V.~Stekhanov}\ITEP
\author{M.~Tarka}\Stony
\author{T.~Tolba}\IHEP
\author{R.~Tsang}\Alabama
\author{K.~Twelker}\Stanford
\author{J.-L.~Vuilleumier}\Bern
\author{A.~Waite}\SLAC
\author{J.~Walton}\Illinois
\author{T.~Walton}\CSU
\author{M.~Weber}\Stanford
\author{L.J.~Wen}\IHEP
\author{U.~Wichoski}\Laurentian
\author{J.~Wood}\WIPP
\author{L.~Yang}\Illinois
\author{Y.-R.~Yen}\Drexel
\author{O.Ya.~Zeldovich}\ITEP
\author{J.~Zettlemoyer}\Indiana

\date{\today}


\begin{abstract}
The EXO-200 Collaboration is searching for neutrinoless double beta decay using a liquid xenon (LXe) time projection chamber.  This measurement relies on modeling the transport of charge deposits produced by interactions in the LXe to allow discrimination between signal and background events.  Here we present measurements of the transverse diffusion constant and drift velocity of electrons at drift fields between 20~V/cm and 615~V/cm using EXO-200 data.  At the operating field of 380~V/cm EXO-200 measures a drift velocity of 1.705$_{-0.010}^{+0.014}$~mm/$\mu$s and a transverse diffusion coefficient of 55$\pm$4~cm$^2$/s.
\end{abstract}

\maketitle

\section{Introduction}
\label{sec-intro}

The EXO-200 experiment uses a liquid xenon (LXe) time projection chamber (TPC) to search for neutrinoless double beta decay (\znbb) of $^{136}$Xe.  Observation of this lepton number violating process would indicate that neutrinos are Majorana particles and could constrain the absolute neutrino mass scale~\cite{Dell'Oro:2016dbc}.  Detection of \znbb, which has a half life in excess of 10$^{25}$~years, requires precise measurement of the energy of the electrons produced in the decay as well as the elimination of sources of background radiation that could obscure a signal.  By employing a TPC, EXO-200 can discriminate between a \znbb signal consisting of two spatially unresolved electrons and $\gamma$ backgrounds that at the energies of interest tend to produce multiple position-resolved energy deposits from Compton scattering.  

Such classification can be improved by a detailed understanding of the diffusion process for electrons as they drift under the influence of the electric field in LXe.  Diffusion effects that are modest in EXO-200, which has a relatively short drift length ($\lesssim 20$~cm), become more important in the ton-scale LXe detectors being developed for both \znbb and dark matter searches, which will have drift lengths $\sim$1~m~\cite{Pocar15,Akerib15cja}. Accurate measurements of the electron diffusion in LXe are therefore important for understanding background discrimination in these next-generation experiments.     

Electron diffusion in LXe has previously been studied at fields higher than those of EXO-200. Early work using short drift lengths ($\lesssim$5~mm) measured the transverse and longitudinal diffusion coefficients for drift fields between $700\ \mathrm{V/cm} < E_{d} <  7500$~V/cm~\cite{Doke:1982Diff,AprileDoke:2010Diff}. In addition, a recent measurement using data from the XENON10 detector determined the longitudinal diffusion coefficient at $E_{d} = 730$~V/cm~\cite{Sorensen:2011diff}. EXO-200 typically operates with a bulk electric field of \Efield, allowing the measurement of the transverse electron diffusion at lower electric fields.

\section{Model for Diffusion in Liquid Xenon}
\label{sec:diff_model}

\begin{figure*}
  \includegraphics[width=0.7\textwidth]{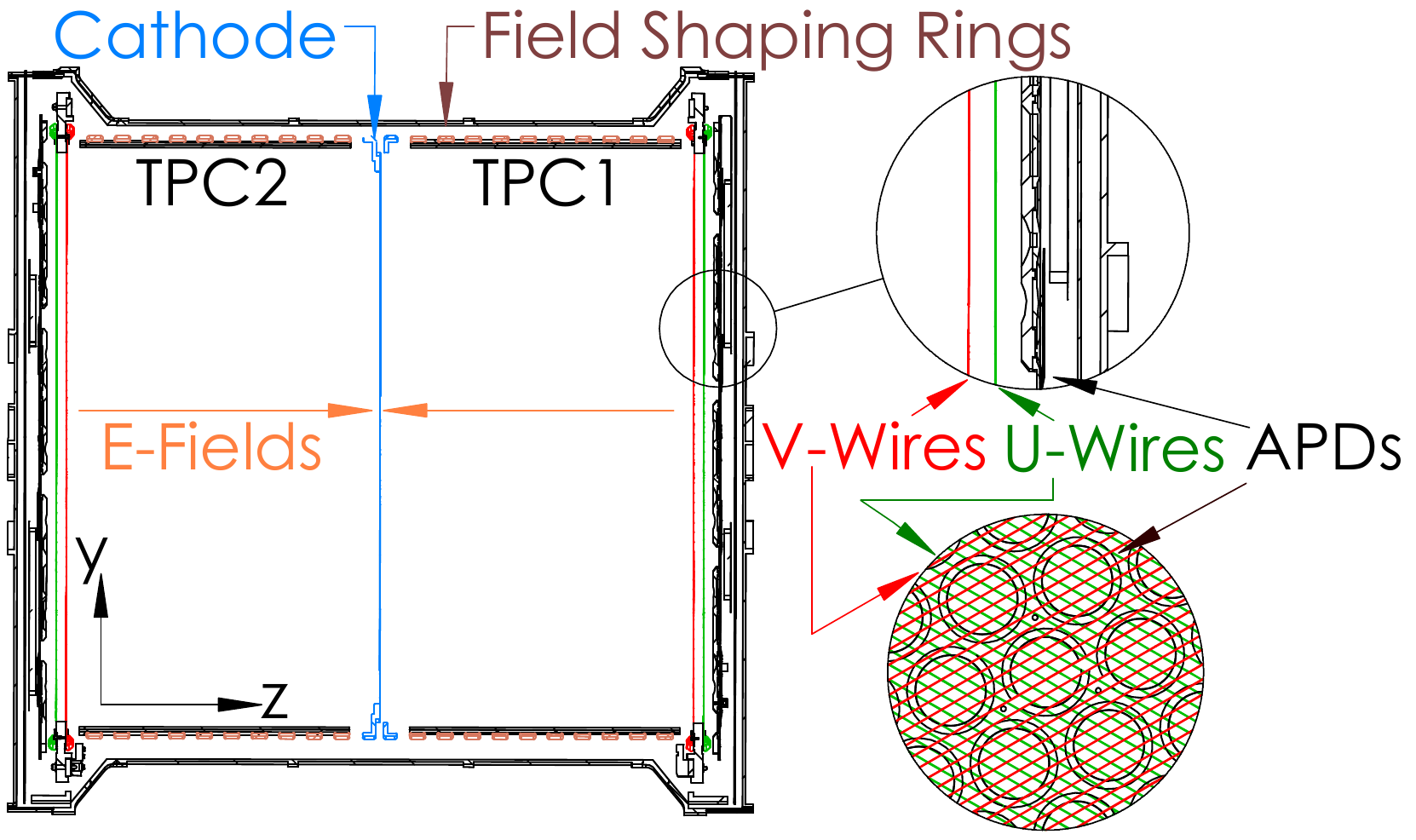}
  \caption{Cut out schematic of the EXO-200 LXe TPC. Shown in the center is the shared cathode plane (blue) and on either side the APD (black), V-wire (red) and U-wire (green) planes are shown. Also included on either side of the cathode are the copper field shaping rings and their support structure.  The top inset shows a zoomed view of the wire and APD planes while the bottom inset shows the same view in the X/Y plane to illustrate the relative angle between the wire planes.\label{EXO_Det}}
\end{figure*}

For an initial $\delta$-function charge deposit of N electrons centered at position $\vec{x} = (0,0,0)$ at time $t = 0$, the charge density, $n(\vec{x}, t)$, at later time $t$ and position $\vec{x}$ can be determined by solving the 3-dimensional diffusion equation~\cite{HUX_Diff}:
\begin{equation}
\begin{split}
n(\vec{x}, t) = \frac{N}{4 \pi D_{T} t \sqrt{4 \pi D_{L} t}}
\exp\left[\frac{-(x^2 + y^2)}{4D_{T}t}\right] \\
 \times \exp \left[ \frac{-(z - v_{d} t)^2}{4D_{L}t} \right]
\label{Diff_Sol_Full}
\end{split}
\end{equation}
where $D_{T}$ ($D_{L}$) is the "transverse (longitudinal) diffusion coefficient", describing diffusion in the directions perpendicular (parallel) to the electric field.  
Equation~\ref{Diff_Sol_Full} describes a Gaussian charge distribution that diffuses while drifting in the $+Z$-direction with velocity $v_d$ due to an applied electric field as shown in Fig.~\ref{EXO_Det}. Different conventions for defining $D_{T}$ exist in the literature; here we use the same convention as in previous measurements of $D_T$ for liquid noble gases~\cite{Doke:1978ArgDiff} where $\sigma^2 = 2Dt$ in each degree of freedom.  Because each degree of freedom can be treated independently, the transverse diffusion process can be modeled as a random walk along the $X$-axis and $Y$-axis where at each time step $dt$ a random step is sampled from independent Gaussian distributions with variance $\sigma^2 = 2D_{T}dt$. 
Considering only the distribution in the 2-dimensional plane transverse to the electric field for an initial point like distribution at $t = 0$ gives a radial variance at time $t$:
\begin{equation}
\langle R(t)^{2} \rangle = \langle x(t)^{2}\rangle + \langle y(t)^{2} \rangle  = 4D_{T}t 
\end{equation}

\section{EXO-200 Detector}
\subsection{Detector description}
\label{sec:Det}
The EXO-200 detector and event reconstruction have been described in detail elsewhere~\cite{Auger:2012gs,Albert:2013_2nuPRC}. The following brief summary focuses on the details relevant to the diffusion measurement.   

The EXO-200 detector consists of a radiopure copper vessel filled with LXe, enriched to $80.672 \pm 0.014\%$ in the isotope $^{136}$Xe~\cite{Albert:2013_2nuPRC}, in which the TPC is immersed.  The detector is $\sim$44~cm long and $\sim$40~cm in diameter and includes two identical back-to-back drift regions known as TPC1 and TPC2 that share a cathode located at the center of the detector.  Charge is detected at either end by a pair of wire planes, each of which lies in front of an array of large-area avalanche photodiodes (APDs)~\cite{Neilson:2009kf}  that are used to observe scintillation light.  The front-most wire plane in each pair (V-wires) is located 192.4~$\pm$~0.5~mm from the cathode and serves both as a shielding grid and to detect induced signals as electrons are drifted from the interaction location to the second wire plane (U-wires) where charge is collected.  The U-wires are then separated from the V-wires by \UToV.  Each APD plane is \UToV behind the U-plane.  The U and V grids at either anode are oriented at a 60$^{\circ}$ angle with respect to each other.  Each grid consists of 114 wires separated by 3~mm, which are grouped into readout channels containing 3 wires each.  This grouping gives 38 readout channels per wire plane that are each 9~mm in pitch.  A cutout view of the detector can be seen in Fig.~\ref{EXO_Det}, which shows the cathode, APD plane and wire planes.

\begin{figure}
\subfloat{\includegraphics[width=1.0\linewidth]{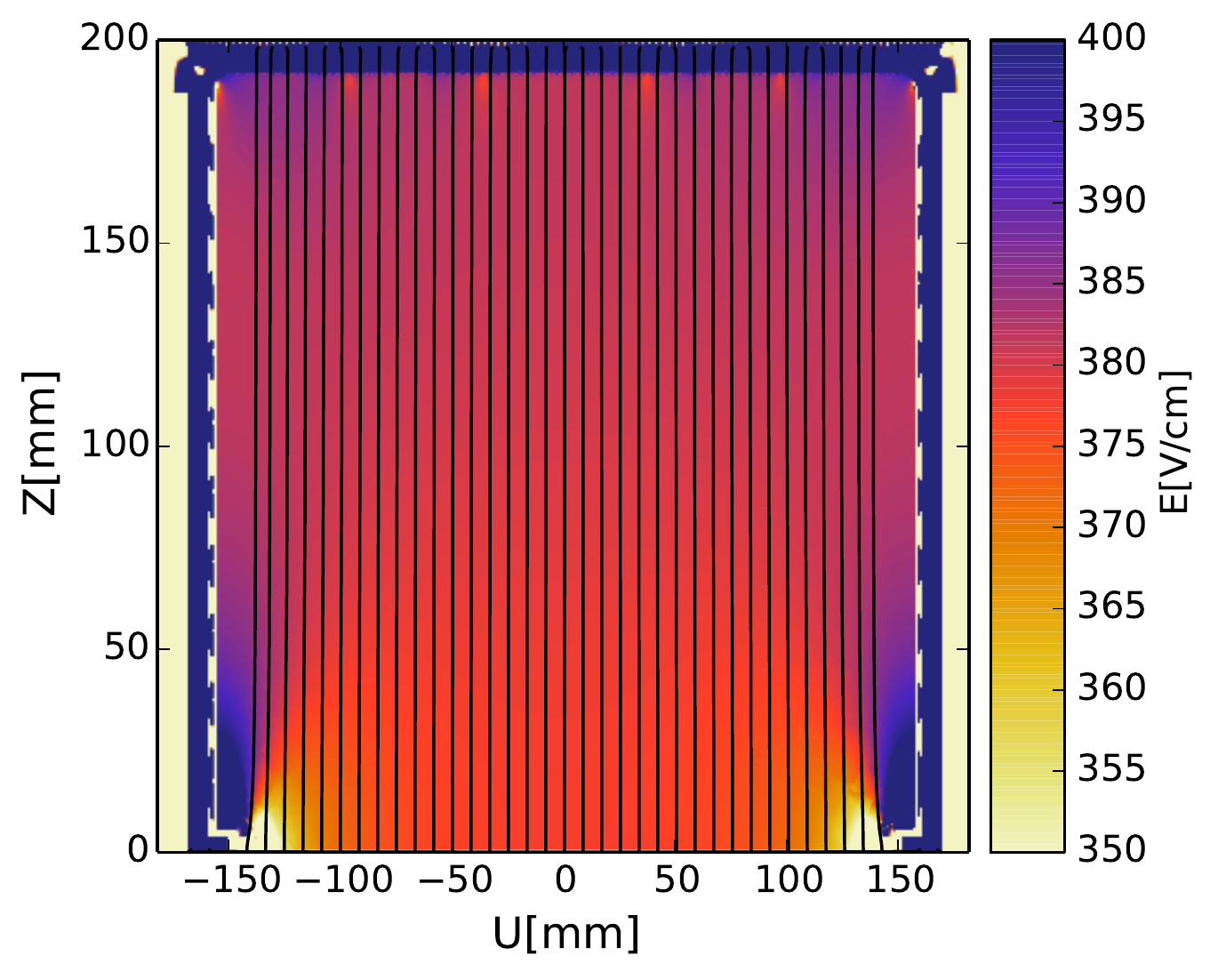}}

\subfloat{\includegraphics[width=1.0\linewidth]{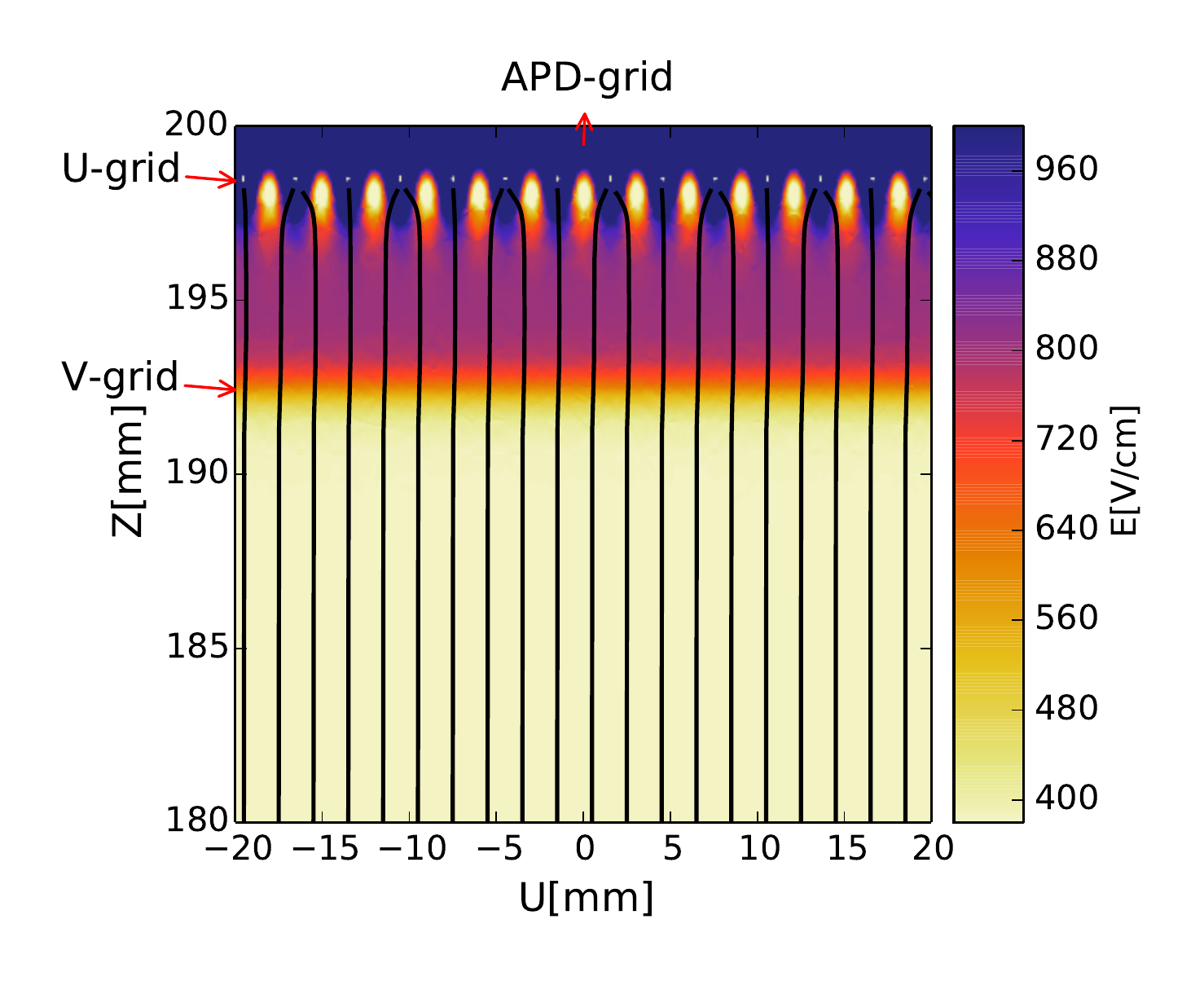}}
\caption{Magnitude of the electric field  in the U-Z plane for (top) one entire drift space of the TPC (cathode is at Z=0~mm) and (bottom) a small segment near the anode.  Also shown in black are the electric field lines that electrons would follow to the collection plane assuming no diffusion. \label{Efield}}
\end{figure}

Particles interacting with the LXe deposit energy by producing both scintillation light (178 nm) and electron-ion pairs (ionization).  Electrons are drifted from their initial location toward the anode by a uniform bulk electric field of \EfieldNoUnit~$\pm$~20~V/cm.  This field is set by holding the cathode at a potential of -8~kV, the V-wire grid at -780~V, the U-wire grid at ground and the APD plane at $\sim-1400$~V.  These potentials are chosen so that the average of the non-uniform field between the U- and V-wire grids (778 V/cm) is approximately twice the main drift field, ensuring full transparency of the V-wire grid and full collection of the U-wire grid for ionization. The magnitude of the electric field in both regions is determined using an electrostatic simulation of the full EXO-200 detector performed with COMSOL~\cite{COMSOL}, and the error on the bulk field denotes the maximum spatial variations around the mean field.  Figure~\ref{Efield} shows a 2D segment of the resulting COMSOL model.  At the edge of each TPC the field is graded in 10 steps by copper field shaping rings to produce a more uniform field along the z-axis within the LXe bulk.  In addition to the data acquired at the standard operating field above, EXO-200 has also acquired smaller amounts of data at fields ranging from 20~V/cm to 615~V/cm.  When operating at these additional fields, the bias applied to the V-wire grid and cathode was chosen to maintain a constant factor of 2 higher mean field in the collection region between the U- and V-wires relative to the bulk of the detector, in order to maintain full transparency of the V-wire grid to electrons.  

Calibration of the detector is periodically performed by positioning $\gamma$-ray sources at several locations around the detector.  Currently four sources ($^{137}$Cs, $^{60}$Co, $^{228}$Th and $^{226}$Ra) are used to span the energy range of interest for the double beta decay search.  These sources can be repeatedly deployed to fixed locations in a copper guide tube that wraps around the outside of the LXe TPC.  For this study only two calibration source locations were used: one centered on the Z-axis behind one of the anode planes and another positioned at the cathode edge (Z=0) and on the +X-axis.

Electrons in LXe can capture on electronegative impurities as they drift, which attenuates the charge signal. To minimize this attenuation, the xenon is continuously circulated through purifiers~\cite{Auger:2012gs}. The purity of the LXe is monitored by periodically measuring the electron lifetime so that a drift-time dependent correction to the event energy can be implemented in the data analysis~\cite{Albert:2013_2nuPRC}.  To limit effects of poor purity, only data for which the electron lifetime is $>$ 2~ms is used in this analysis.  

\subsection{Monte Carlo simulations}
\label{sec:MC} 

Previous EXO-200 analyses have relied on Monte Carlo (MC) simulations to understand the detector response to ionizing radiation. These MC simulations are described in detail in previous papers~\cite{Albert:2013_2nuPRC} and are validated using data from the $\gamma$ source calibrations. In these analyses, it was not necessary to include electron diffusion given its small effect ($\sim$1~mm over the full drift) relative to the 9~mm channel size.  However, more recent efforts to optimize background discrimination have focused on accurately modeling the size and multiplicity of charge deposits.  This requires incorporating diffusion into the EXO-200 MC simulation.  

For a given source the MC simulation first produces energy deposits within the detector using a GEANT4-based application~\cite{G4Paper} which employs a detailed model of the detector geometry~\cite{Albert:2013_2nuPRC}.  The ionization deposits are binned into cubic voxels with 0.2~$\mu$m edges.  Each voxel is then tracked as it is drifted from the interaction location to the collection wires using a simulation of the electric field in the TPC.  For every time step, the charge induced on each readout channel is determined from the Shockley-Ramo theorem~\citep{Shock,Ramo:1939vr}.

To model diffusion, at each time step of size $dt$, additional random displacements, $dx$ and $dy$, are added to the position in the $x$ and $y$ direction.  As described in Sec.~\ref{sec:diff_model}, these displacements are drawn from independent Gaussian distributions with $\sigma^2 = 2 D_T dt$.  To ensure sufficient granularity of the charge distribution, the initial 0.2~$\mu$m voxels are further split into equal deposits consisting of a maximum of $\sim$600 electrons prior to simulating their drift.  MC studies indicated that this level of granularity is sufficient to accurately model the diffusion process, and the resulting reconstructed charge deposits are not affected by finer pixelization.

In previous publications~\cite{Albert:2013_2nuPRC}, the signal generation stage used a two-dimensional (2D) electrostatic modeling of the electric field in the detector and assumed azimuthal symmetry.  This approximation does not accurately reflect the 60$^{\circ}$ orientation of the U-wires relative to the V-wires, which has a substantial effect on the trajectory of the charge deposits in the collection region.  After incorporating diffusion, this 2D simulation gave poor agreement to data in the multiplicity and amplitude of signals induced on U-wires neighboring the collection channel.  Implementing a full three-dimensional (3D) simulation of the electric field substantially improved the agreement of these signals between data and simulation.  To optimize the speed of the simulation, only a segment spanning 4mm x 4mm x 25mm, with edges oriented parallel to the U- and V- wires, was used to approximate the field throughout the detector, assuming translational symmetry.  This does not fully account for imperfections in the electric field near the edge of the detector, but provides a good description within the fiducial volume defined by the hexagonal region with 162~mm apothem that is $>$10~mm from the cathode and each V-wire plane.~\cite{Albert:2014awa}  For data taken at lower fields, a deficit of events at high radius near the cathode indicated that significant radial non-uniformities were present in the drift field.  To avoid introducing systematic errors due to these non-uniform regions, the radial fiducial volume cut was tightened to remove these regions from the analysis.  As described in Section~\ref{sec:diff_fits}, checks were performed to ensure no radial dependence was present within the fiducial volume used at each field.
  
A charge propagation simulation is used to produce waveforms for each readout channel with the same parameters as those obtained in data.  Each waveform is sampled at 1~MHz and contains 2048 samples centered around the interaction time. The pulse amplitudes are determined using the gain of the detector readout electronics in ADC counts per electron charge.  Noise traces acquired with the detector throughout its operation are added to the simulated waveforms to accurately model the detector noise. Waveforms for both MC and data are processed with the same analysis code to reconstruct signals on the U- and V-wires.

The U- and V-wire signals reconstructed from the data and MC waveforms are grouped into "clusters" of signals arising from the same interaction, based on the position and timing information of all signals in an event.  For each cluster, the total energy, position, and the number of active readout channels are determined. The number of active channels in each cluster depends on the initial deposit size and the transverse diffusion as the deposit is drifted to the collection grid.  The distribution of the number of wire channels contributing to clusters can be used to measure the effects of diffusion in the EXO-200 data, as will be described in Sec.~\ref{sec:diff_fits}. 

\section{Results}

\subsection{Drift velocity}

In order to measure the diffusion constant, the transport of electrons in LXe must be accurately modeled.  The simulation described in Section~\ref{sec:MC} requires a realistic electron drift velocity to produce signals with the correct shape and timing structure.  Previous works have reported the drift velocity measured by EXO-200 at the standard operating field of \Efield~\cite{Albert:2013_2nuPRC}.  Here, the velocity is measured over the larger range of electric fields described in Section~\ref{sec:diff_fits}.  

The drift velocity is determined from the drift time for events originating from a known location in the detector.  To perform the measurement in the region of the detector between the cathode and the V-wire plane (i.e., the ``bulk velocity''), events that drift the full length of the TPC are used.  The drift time is converted to an average velocity using the separation between the cathode and V-wire grid of \CathToV.  Two separate analysis techniques were used to check for consistency.        

The first analysis measures the maximum drift time for $\alpha$ events emitted from surface contamination on the cathode and V-wire plane.  Due to the short range of an $\alpha$ in LXe, the separation between the initial charge distribution and the originating surface is small relative to the uncertainty in the reconstructed Z-position~\cite{Albert:2013_2nuPRC}. Energy deposits from $\alpha$ events have higher charge density relative to $\gamma$s and $\beta$s causing substantially more scintillation from recombination relative to ionization.  This allows $\alpha$ events to be selected using a cut on the ratio of scintillation light to charge yield which is tuned using peaks from known $\alpha$ backgrounds~\cite{Albert:2015vma}.  To obtain sufficient statistics for this measurement requires several days of "low background" data, for which a calibration source is not present.  In 580~days of low background data taken at the standard field of \Efield, $\sim$100k $\alpha$ events were observed.  In addition $\sim$1k $\alpha$ events over 4~days of data taking at 567~V/cm were acquired.  Low background data were not taken at any of the other fields considered here.    

\begin{figure}
    \includegraphics[width=1.0\linewidth]{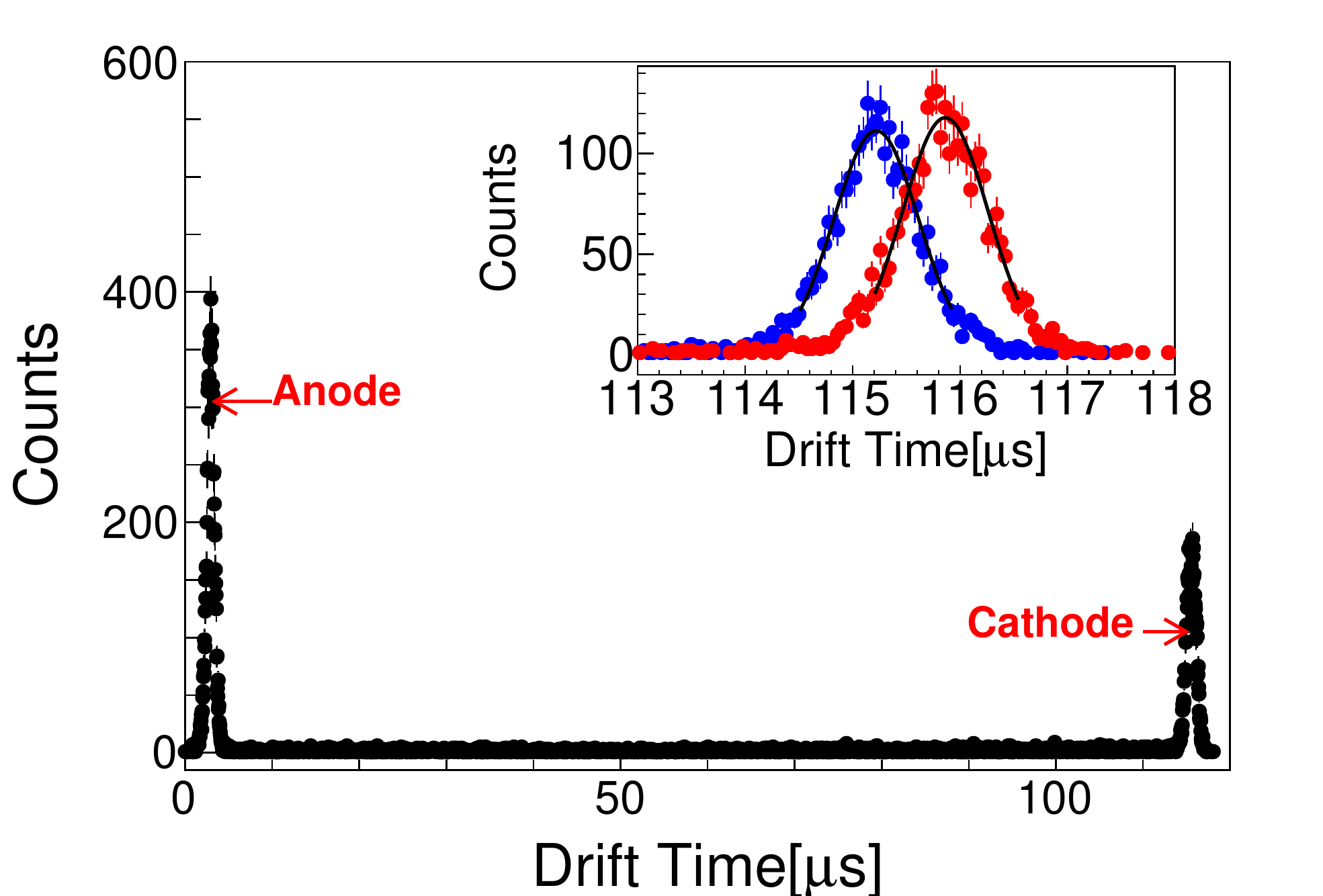}
     \caption{Combined distribution of drift times for $\alpha$ events from both drift spaces.  The inset shows events near the cathode separately for TPC1 (blue) and TPC2 (red).  Also included are the best Gaussian fits used for determining the maximum drift time in both TPCs. The offset between the TPCs corresponds to a $\sim$0.5~mm uncertainty in the cathode position as explained in the text.}
    \label{DVCalc:Alpha}
\end{figure}

As shown in Fig.~\ref{DVCalc:Alpha}, the distribution of drift times from $\alpha$s has a large concentration of events arising from surface contamination on both the cathode and V-wire planes. The distribution of drift times from both surfaces are separately fit to Gaussians to determine the mean drift time for events originating on each.  The electric field is less uniform between the U- and V-wire planes than in the bulk, as shown in Fig.~\ref{Efield}. For this reason it is not possible to define a single drift velocity in this region, but the mean drift time between the U- and V-wire planes can still be determined and is subtracted in the measurement of the bulk drift velocity.   

The dominant systematic with this technique is due to the measurement of the distance between the cathode and the V-wire plane.  This separation is determined using the known geometry at room temperature as well as the expected thermal contraction of the TPC materials at LXe temperatures.  After cooling the TPC, there is some uncertainty on the resulting drift length and the position of the cathode relative to the wire planes.  As shown in the inset of Fig.~\ref{DVCalc:Alpha}, a $\sim$$1\ \mu$s difference in the drift time is seen between the drift spaces of TPC1 and TPC2 for $\alpha$ events.  This difference is consistent with an offset of the cathode position from center by $\sim$0.5~mm.  This potential offset is taken as an additional systematic error on the total distance between the V-wire plane and cathode in each TPC.

\begin{figure}
    \includegraphics[width=1.0\linewidth]{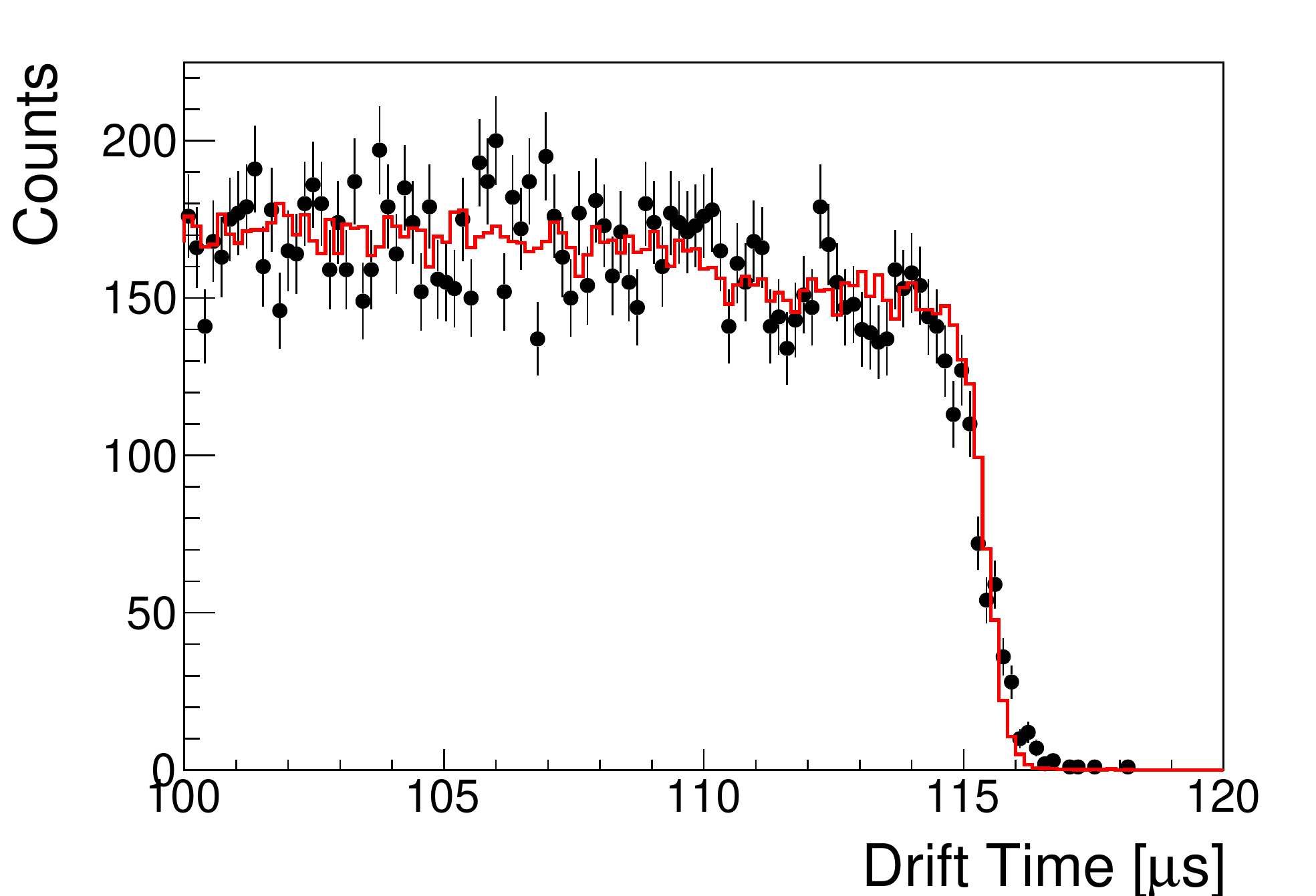}
    \caption{Distribution of drift times for events with the longest drift times for MC (red line) and data (black points) for a $^{228}$Th source positioned near the cathode.}
    \label{DVCalc:Th}
\end{figure}

In addition to using $\alpha$ events from surface contamination, $\gamma$-ray source calibration data can be used to measure the drift time. Data taken with a $^{228}$Th source at the cathode was used in order to maximize the number of events traveling the full length of the detector.  Due to the finite size of charge deposits and position-dependent variations in the drift length, the distribution of drift times has a finite width edge at its maximum, as shown in Fig.~\ref{DVCalc:Th}.  A simulation is used to calculate the position along this edge that corresponds to the average drift time for events traveling the full length of the TPC in MC.  

Following the same procedure as for the measurement using $\alpha$ events, the portion of the total drift time corresponding to the transit time between the V- and U-wires must be subtracted to obtain the drift time in the main drift region alone.  This V-to-U drift time is directly measured using $\alpha$ events at the two fields for which low background data were acquired (380 V/cm and 567 V/cm) as described above.  For the remaining fields, this drift time is estimated using the known spacing between the V- and U-wires and the measured value of the drift velocity at the mean field in this region.  Since the field in the U-V gap was chosen to be twice the field in the main drift region for each dataset, and since the field was also varied by a factor of two between different data sets, the measurement of the drift velocity in the bulk region for each higher field point can be used to estimate the drift velocity in the collection region at the point below it.  While this approximation neglects the non-uniformity of the electric field between the U- and V- planes, the correction itself is small, and the resulting error gives a negligible contribution to the total error.

From Fig.~\ref{DVCalc:Th} it can be seen that drift time distribution in data are not fully reproduced by MC.  In particular, the tail of the distribution in data are slightly broader than in the simulated distribution. This difference may arise from the approximate modeling of the electric field geometry near the TPC surfaces in simulation. To ensure that any differences in the detailed modeling of the edge of the distribution are accounted for, the width of the edge in the data distribution is added as a systematic error in the measurement of the maximum drift time.

\begin{figure}
   \includegraphics[width=1.0\linewidth]{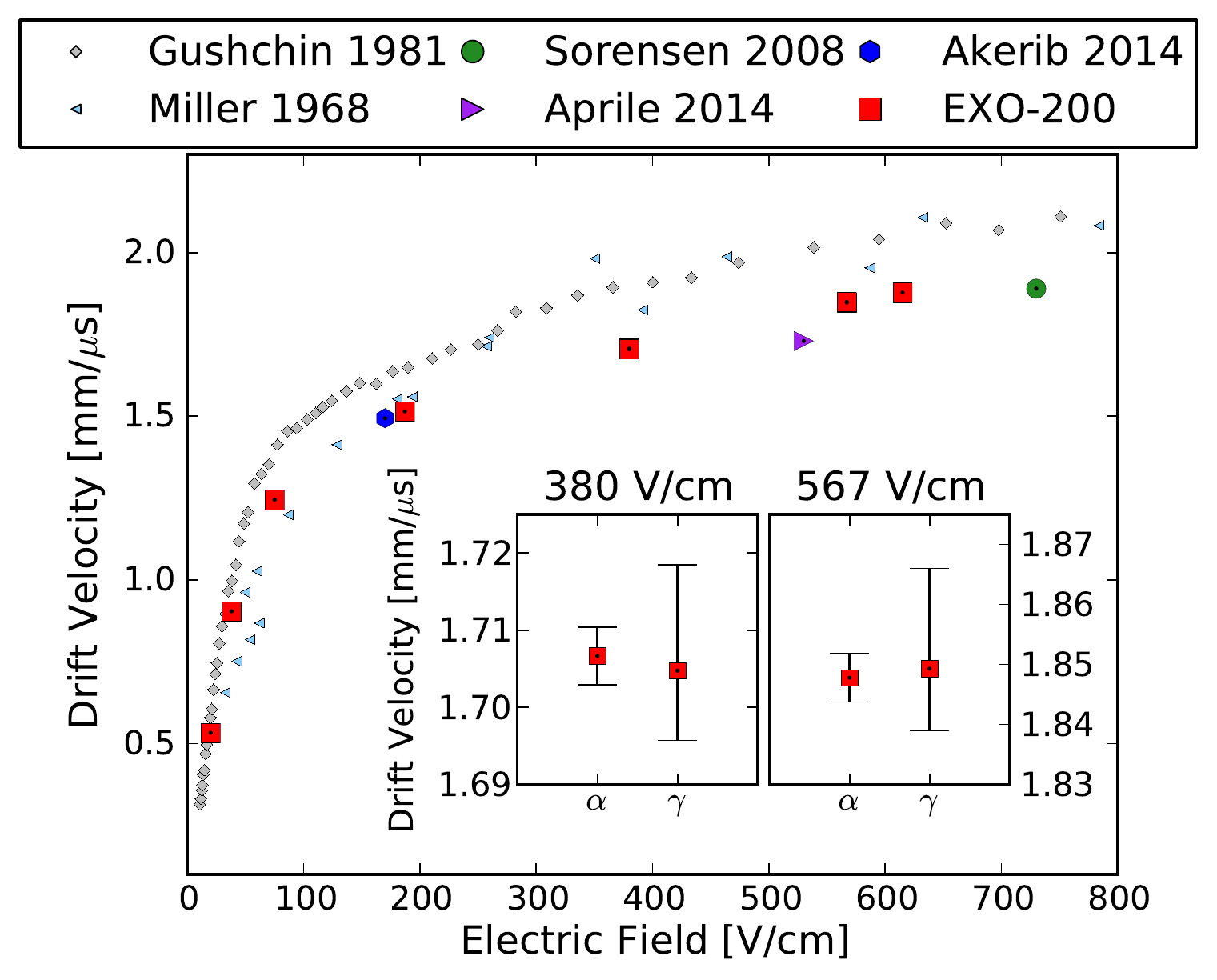}
   \caption{Drift velocity measured using EXO-200 for $\gamma$s in $^{228}$Th source data and $\alpha$s in low background data (167~K, red).  Also included are results from Miller et al. (163~K, cyan) (Miller 1968:~\cite{Miller:DriftV}), Gushchin et al. (165~K, grey) (Gushchin 1981:~\cite{Gushchin_DV}) and the XENON10 (177~K, green) (Sorensen 2008:~\cite{Sorensen:2008}), XENON100 (182~K, purple) (Aprile 2014:~\cite{Aprile_X100_DV}) and LUX (175~K, blue) (Akerib 2014:~\cite{LUX_DV}) collaborations.  LXe operating temperatures are included to account for possible variations due to temperature~\cite{Benetti_DVTemp}.  The inset shows a comparison of the two methods of measurement at 380~V/cm and 567~V/cm in EXO-200.}
   \label{EXO_DV}
\end{figure}

The measured drift velocity from $\alpha$ events and source data at each field are shown in Fig.~\ref{EXO_DV}.  Both techniques agree within errors at the fields at which data are available for both (inset).  Figure~\ref{EXO_DV} also shows previous measurements of the drift velocity of electrons in LXe~\cite{Miller:DriftV,Gushchin_DV,Sorensen:2008,Aprile_X100_DV,LUX_DV}. The EXO-200 measurements differ from previous measurements at the same fields by up to $\sim$10$\%$. The difference between the EXO-200 measurements and the early measurements by Miller et al.~\cite{Miller:DriftV} and Gushchin et al.~\cite{Gushchin_DV}, using substantially shorter drift distances, cannot be fully explained by the temperature differences between the measurements.  Previous measurements have found the drift velocity to vary by $\sim$~0.012~mm/$\mu$s~K$^{-1}$ at similar fields~\cite{Benetti_DVTemp}.  In the current measurement, data were taken at $167.0 \pm 0.2$~K, while the Miller et al.~\cite{Miller:DriftV} and Gushchin et al.~\cite{Gushchin_DV} measurements were obtained at lower temperatures (163~K and 165~K respectively).  These temperature differences are consistent with slower drift velocity at higher temperatures. However, they would account for only a $\lesssim 3\%$ variation between the measurements shown in Fig.~\ref{EXO_DV}.  The more recent measurements from XENON10~\cite{Sorensen:2008}, XENON100~\cite{Aprile_X100_DV} and LUX~\cite{LUX_DV} are taken at 177~K, 182~K and 174~K respectively.  These measurements take advantage of the $\gtrsim$10~cm drift lengths and higher chemical purity available with modern LXe TPCs and are in better agreement with the EXO-200 measurements presented here.  The residual disagreement between EXO-200 and these more recent measurements is comparable to the expected deviation due to temperature differences.  

In addition, early studies suggested that the purity of the LXe can affect the drift velocity of electrons in LXe~\cite{Yoshino:1976zz,SpearPurity}.  Although only high purity data were included for the measurement shown in Fig.~\ref{EXO_DV}, the drift velocity was also measured for data with lower purity. For this data electron lifetimes varied between 80~$\mu$s and 600~$\mu$s.  The drift velocity for this data were measured using $\alpha$ events following the method outlined above and agreed within error to the measurement at purity $>$2~ms included in Fig.~\ref{EXO_DV}.

\subsection{Diffusion}
\label{sec:diff_fits}
At the standard operating field of \Efield, data were taken using $^{226}$Ra, $^{228}$Th and $^{60}$Co calibration sources, as described in Section~\ref{sec:Det}.  Runs were performed with the sources positioned near both the anode and cathode to account for any systematic errors that may arise from different position and energy distributions.  Variations in the measured diffusion coefficient for different data sets were at most $\sim$10$\%$.  For data taken at operating fields other than \Efield, only the $^{228}$Th source was used.  A minimum of $\sim$100k events were used for the measurement at each field.

\begin{figure}
\subfloat[$^{228}$Th]{\includegraphics[height=0.5\linewidth, width=1.0\linewidth]{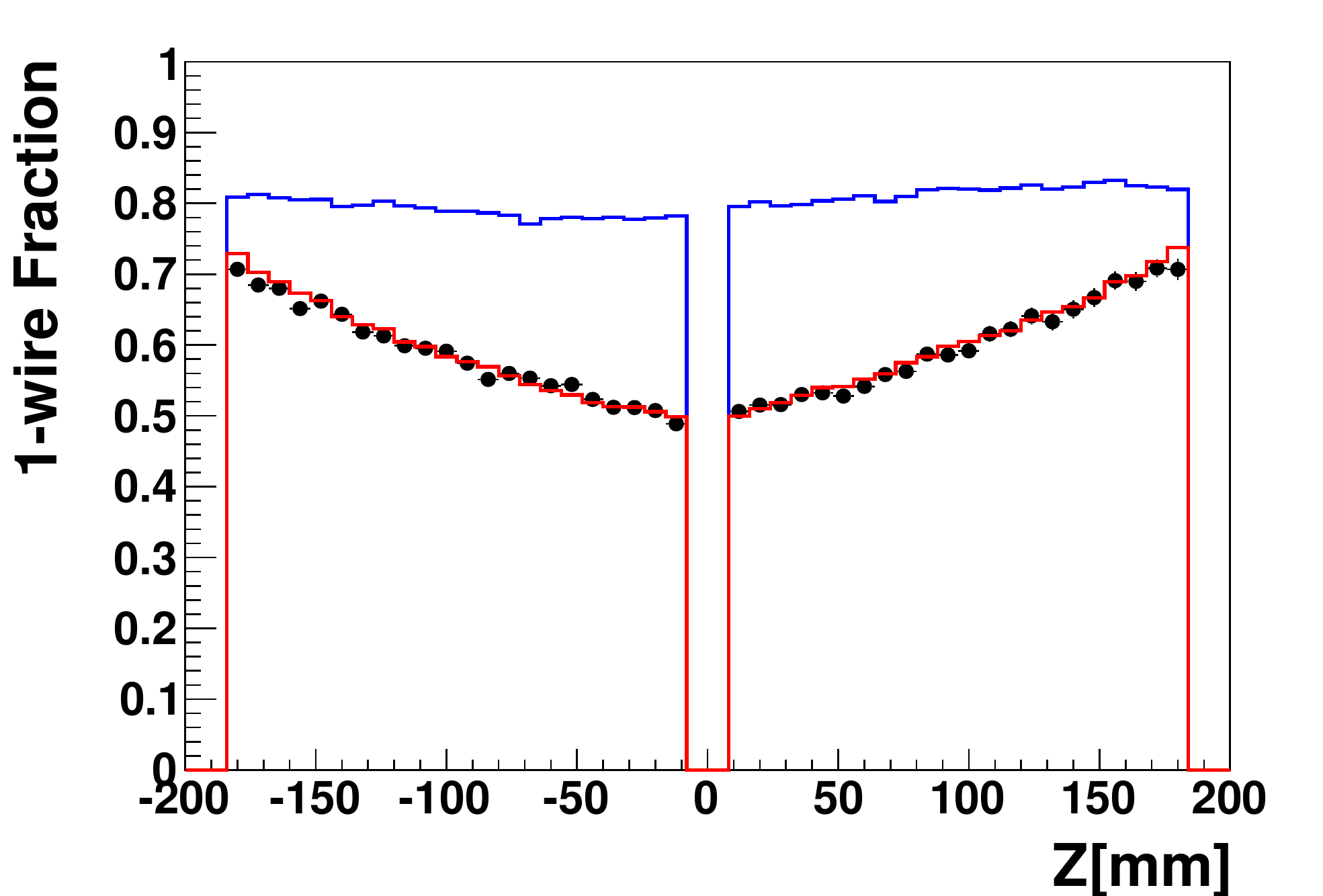}\label{MCEffect:Th}}

\subfloat[$^{60}$Co]{\includegraphics[height=0.5\linewidth, width=1.0\linewidth]{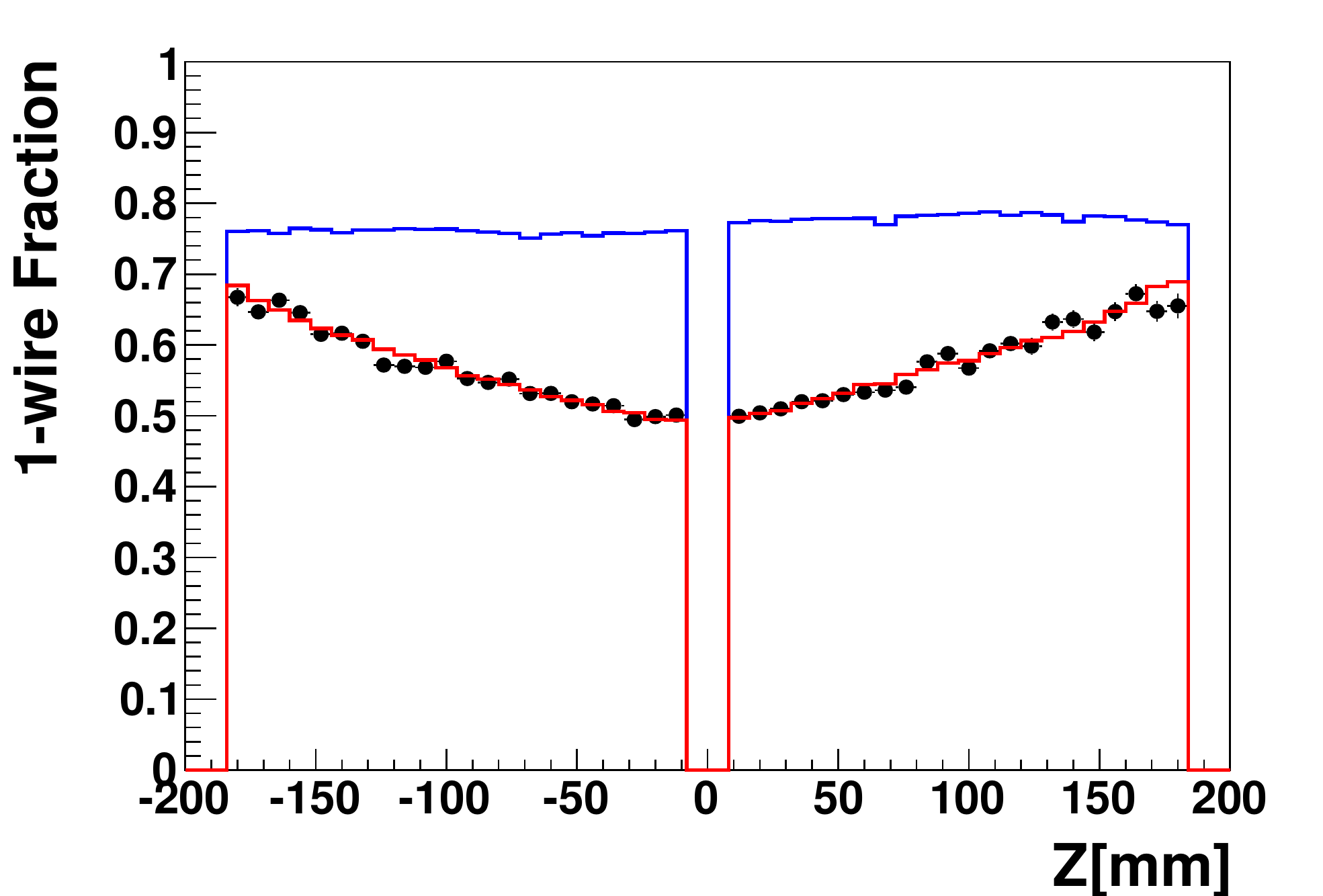}}

\subfloat[$^{226}$Ra]{\includegraphics[height=0.5\linewidth, width=1.0\linewidth]{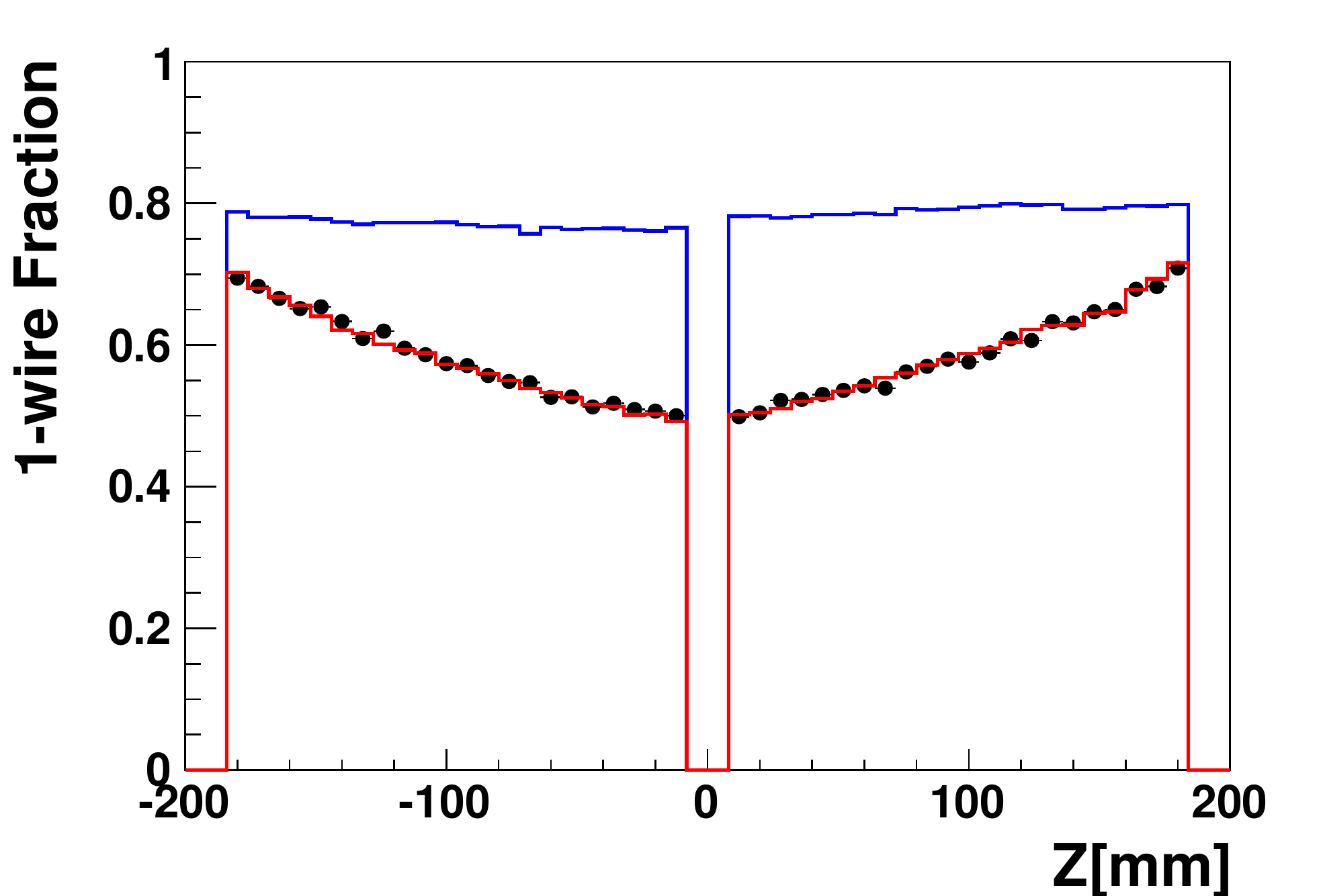}}
  
\caption{EXO-200 comparison of 1-wire fraction between Data (black points) and MC without diffusion (blue line) and with diffusion (red line) for operation at \Efield.  The sources shown are (a) $^{228}$Th,  (b)$^{60}$Co and (c) $^{226}$Ra.  For MC with diffusion a coefficient at the best fit of 55~cm$^2$/s was used for all sources.}
\label{MCEffect}
\end{figure}

MC was generated using a range of diffusion coefficients.  For each simulated coefficient, the distribution of the fraction of clusters containing only 1 readout channel instead of 2 readout channels (i.e., the ``1-wire fraction'') for different $Z$-positions was determined and compared to that observed in data.  A maximum likelihood fit is performed to determine the diffusion coefficient that gives the best agreement between data and MC.  The distributions of 1- and 2-wire events for each source are binned into 6 $Z$-bins spanning the length of both TPCs. Only events inside the fiducial volume described in Section~\ref{sec:MC} are used to avoid field non-uniformity near the edges of the detector.  Fits are performed separately for each source and source position, and results from the fit to each dataset agree within errors.  The final measurement at each field is then performed using a combined fit to all sources.  

Figure~\ref{MCEffect} shows the $Z$-dependence of the 1-wire fraction for data and MC for each of the three sources at the standard field.  This includes MC with a diffusion coefficient near the best fit value of \textit{$D_{T}$}~=~55$\pm$4~cm$^2$/s as well as that generated without diffusion.  A linear dependence in the 1-wire fraction versus $Z$-position can be seen in source calibration data, consistent with increasing event size at longer drift lengths. For the MC without diffusion, the distribution does not depend on drift length.  As shown in Fig.~\ref{MCEffect}, incorporating diffusion into the MC results in good agreement in the distribution of the 1-wire fraction with data.

The simulation of the EXO-200 detector response has been studied in detail to optimize its agreement with data, but some differences remain.  Inaccuracies in the simulation of 1- and 2-wire event distributions are estimated by measuring the differences in these distributions between data and MC for events within 10~mm of the V-wire plane, where the effect of diffusion is negligible.  Differences in the 1-wire fraction for each $Z$-bin between 5-20$\%$ was observed at the fields considered here. The largest of these errors was associated with the fields farthest from the standard operating field, where the MC is less studied.

To account for the error associated with systematic difference between the data and MC, an additional scaling of the 1-wire and 2-wire event distributions in each bin is allowed to vary.  These scalings are constrained in the likelihood fit by Gaussian errors determined from the differences between data and MC for the short drift time events.  An additional independent scaling of the overall ratio between 1-wire and 2-wire events was also included and constrained using the same technique.  These scalings and constraints allow the effect of any errors in the MC simulation of the expected wire distributions to be propagated as systematic errors to the measurement of the diffusion constant.  No correlation between the constraints were observed so the constraints were separately fit for each data set and the likelihood was summed over all data sets to determine the best fit diffusion constant and error at each field.

In addition, there is an error associated with inaccurately reproducing the signal amplitude of charge channels in MC. At the standard field of 380~V/cm, the relative signal amplitude of each channel is empirically scaled to match the amplitude observed in data to within $10\%$. At lower electric fields, the signal amplitude is reduced due to the lower charge extraction efficiency.  This variation in charge yield with electric field has been well studied in previous experiments~\cite{Conti:2003av,AprileDoke:2010Diff}. The relative summed signal amplitude is measured for events in the $^{228}$Th peak for each field and a correction to the expected charge yield in simulation is applied by scaling all charge signals by the observed factor.  This does not fully account for the detailed microphysics of this process, and future work could use simulation packages such as NEST~\cite{NEST_SIM,G4Paper} to account for the varying charge yield with field on an event-by-event basis.  Since only an empirical scaling is used here, an additional systematic error is estimated by performing the fit to the diffusion constant for both the scaled and un-scaled U-wire signals. The shift in the central value of the fit is then included as an additional systematic error to conservatively account for any error in the simulated charge yield with field.  The resulting error from the signal magnitude systematic is $5\%$ at the standard field compared to the $8\%$ total error.  For the measurements at lower fields, this error increases due to the larger variations in charge yield. For the lowest field, this systematic results in a $46\%$ error compared to the $48\%$ total error.

The effect of self-repulsion of charge clouds is ignored in the full simulation of charge propagation and diffusion. This effect was studied separately to determine its impact on the measurement and is qualitatively different from diffusion since it depends on the density of the charge distribution.  This introduces a dependence on the event energy and an additional non-linear $z$-dependence in the size of the charge distribution versus drift time. However, the variation with cluster energy for the diffusion coefficients measured here is less than $10\%$ over the full energy range and is consistent with errors. This indicates that the data are not sensitive enough to show the effect of self-repulsion of charge deposits.  

As a cross-check, a standalone simulation including the effects of both self-repulsion and diffusion was performed, using a similar technique as in \cite{Doke:1978ArgDiff}.  This simulation tracked the drifting electrons while accounting for both diffusion as well as repulsion from each electron to calculate the expected spread of the initial deposit. The screening of polarized LXe ions was ignored to give a conservative estimate of the error from the repulsion process.   Results of this study confirm that repulsion is a negligible effect, contributing at most $3\%$ to the overall spread of the charge distribution. 

\begin{figure}
  \centering
  \includegraphics[width=1.0\linewidth]{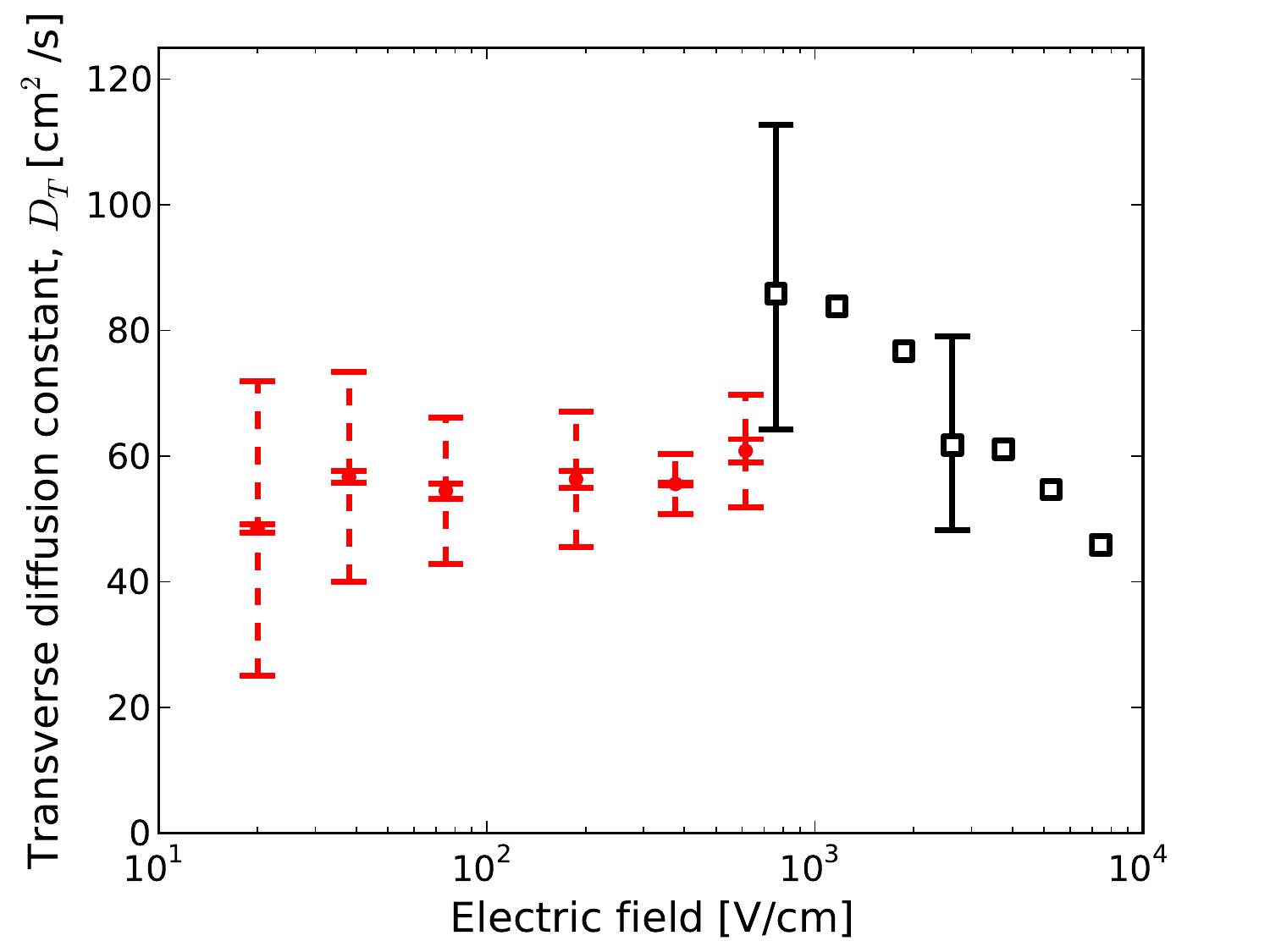}
  \caption{Measured diffusion coefficient versus electric field in EXO-200 (red points).  Solid error bars are statistical and dashed errors include the systematic errors discussed in the text.  Also plotted are previous measurements made by T. Doke and collaborators (black points)~\cite{Doke:1982Diff,Doke:Swarm,Doke:Portugal,AprileDoke:2010Diff}.  Error bars for the previous measurements from~\cite{Doke:Portugal} are shown, where errors were provided only for two of the seven points. }
\label{EXO_Diff}
\end{figure}

As described in Section~\ref{sec:MC}, the fiducial volume at each field was defined to minimize systematic effects from field non-uniformities. To check for any residual systematic effects, the diffusion measurement was compared for different radial segments in the detector. At the standard operating field of 380 V/cm, measurements using high- and low-radius segments of the detector were consistent, and these measurements constrained any residual variation between the segments to $<$~2.6$\%$, compared to the 8$\%$ total error. For the lowest field at 20 V/cm, where the field non-uniformities were greatest, the maximum variation was $<$~5.5$\%$ compared to the 48$\%$ total error. This shows the resulting systematic effect of residual non-uniform fields in the fiducial volume is small relative to the systematic error from other sources

Figure~\ref{EXO_Diff} shows the measured diffusion coefficient from EXO-200 data from the fields used, as well as results of a previous measurement~\cite{Doke:1982Diff,Doke:Swarm,Doke:Portugal,AprileDoke:2010Diff}.  The error bars for the previous measurement are taken from~\cite{Doke:Portugal}, where errors in the measurement of $eD_{T}/\mu$ are shown by the authors for only two of the seven points.  The diffusion coefficient is calculated by the authors using the transverse spread of the electron clouds and the drift velocity measured in~\cite{Doke:Portugal}, assuming the relationship to the mobility $v_{d}$ = $\mu E_{d}$, where $\mu$ is the electron mobility, $v_{d}$ is the electron drift velocity, and $E_{d}$ is the electric field~\cite{Doke_2004}.  The resulting measurements of the diffusion coefficient shown in~\cite{Doke:1982Diff,Doke:Swarm,Doke:Portugal,AprileDoke:2010Diff} do not include the error bars on these two points from the original reference, but they are propagated to the measurement of $D_T$ shown in Fig.~\ref{EXO_Diff} here for comparison.  Although there is no direct overlap in the fields considered, the results of the current study are consistent within errors near 600~V/cm, but suggest a 25$\%$ lower central value than the previous measurement.  
These results use a maximum drift length of 198~mm, which is substantially larger than the 2.7~mm drift length used in the previous measurement~\cite{Doke:Portugal}.  This longer drift length limits the effect of systematic errors on the measurement of the diffusion coefficient.  Finally, the detector MC and event reconstruction available for EXO-200 data allows this analysis to account for systematic effects in the measurement of the charge distribution versus drift length in detail.

\section{Conclusions}

A measurement of the drift velocity, $v_d$, and the transverse diffusion coefficient, $D_T$ for electrons in drift fields between $20~\mathrm{V/cm} < E_d < 615$~V/cm has been performed using data from the EXO-200 detector.  These results contribute to the understanding of the transport properties of electrons in LXe.  Ton-scale LXe particle detectors, which are currently in development for the search for $0\nu\beta\beta$ and dark matter, will be more sensitive to effects such as diffusion because of their longer drift lengths, making this measurement relevant for the development of future LXe technology. 

\begin{acknowledgments}
We would like to thank M. Anthony and E. Aprile for helpful comments on the previous measurements of the diffusion coefficient shown here. We also gratefully acknowledge the KARMEN collaboration for supplying the cosmic-ray veto detectors, and the WIPP for their hospitality. EXO-200 is supported by DOE and NSF in the United States, NSERC in Canada, SNF in Switzerland, IBS in Korea, RFBR in Russia, DFG Cluster of Excellence “Universe” in Germany, and CAS and ISTCP in China. EXO-200 data analysis and simulation uses resources of the National Energy Research Scientific Computing Center (NERSC).
\end{acknowledgments}

\bibliography{references}
\end{document}